\documentstyle[12pt]{article}
\begin{document}
\begin{center}
{\bf Semiclassical treatment of the ground state \\
rotational band and of the isovector angle vibration \\
for deformed superfluid nuclei\footnote{first printed 
as preprint FT-326-(1988)/February, by the Institute for Physics and
Nuclear Engineering, Bucharest, Romania.} 
 }
\\[.5cm]
Marius Grigorescu \\[1cm]
\end{center}
\noindent
{\bf Abstract:} The definition of the pairing interaction for rotating
systems is discussed, and a simple, mainly analytic model for its 
treatment is formulated. The results are used to describe microscopically
the giant angle dipole resonance of the superfluid deformed nuclei in the
semiclassical approximation of the cranking model. \\[1cm]
{\bf I. Introduction}\\[.5cm] \indent
The microscopical description of the superfluid rotating nucleus has been the
object of a constant interest during the time, started after the early 
attempts, made in solid state physics, to treat the similar situation of 
superconductors placed in magnetic fields \cite{1} \cite{2}. This similarity
is better shown in the semiclassical approximation given by the cranking
model, where the angular velocity from the Coriolis term is the 
correspondent of the external magnetic field. For the deformed nuclei, 
initially the main problem was to obtain the quasiparticles of a hamiltonian
consisting of a cranked single-particle Nilsson term, with pairing interaction
\cite{3}. The further studies were concentrated on a more elaborated
treatment of the symmetries, using the Hartree-Fock-Bogolyubov (HFB) solutions
of the cranking hamiltonian in order to achieve new trial wave functions
by angular momentum \cite{4} or particle number \cite{5} \cite{6} projection.
Considering the cranked HFB theory as a large-deformation limit, for a more
accurate treatment by variation after projection \cite{7} \cite{8}, the 
spurious rotational energies were also studied \cite{9}. \\ \indent
Beside the problems raised by the breaking of the symmetries, there is also
another important point, related to the choice of the single-particle basis
used to define the pairing interaction term for systems which do not have
a time-reversal invariant single-particle hamiltonian. As it was remarked
before in the case of the electrons in solids with non-magnetic impurities,
the correct choice is given by the eigenfunctions of the whole single-particle
part of the hamiltonian. This part includes the perturbation terms, and the
final result cannot be reached simply by re-writing the pairing term from the
unperturbed basis to the new one \cite{1}. For nuclei a similar treatment 
was given in \cite{4}, where the pairing interaction defined in the Nilsson
basis is re-written numerically in the basis where the Nilsson plus Coriolis
single-particle term is diagonal. Later on, a different choice was presented
in ref. \cite{5}, and used to estimate the increase in the moment of inertia
due to the Coriolis anti-pairing and rotational alignment phenomena. At
backbending, the quasiparticle vacuum was proved to become unstable, with 
effects similar to the gapless superconductivity already known from the
reference \cite{10}, where the Nilsson basis was used. \\ \indent
In this paper the basis problem is treated in a way similar to the one
presented in \cite{4}, but with two exceptions. First, the pairing 
interaction is defined in the new basis by using the time-reversal
operator, and second the single particle hamiltonian is restricted to the
cranked anisotropic harmonic oscillator (C.O.) term. This term can be
diagonalized analytically \cite{11} \cite{12}, and moreover, the restriction
is able to give a model where the relevant physics of the spin-independent
phenomena combining the effects of deformation, pairing and rotation is
maintained in the simplest way.  \\ \indent
The interest in such a treatment is twofold, because it gives us a better
understanding of the pairing interaction, and also because the flow pattern,
easy to obtain by a cranking calculation \cite{13}-\cite{15}, is important
for the microscopical description of the isovector angle vibrations (A.V.) 
\cite{16}. Even though the picture of two ellipsoids in opposite angular
rotational oscillation, initially proposed for these modes \cite{17}, was
subject to criticism \cite{18} \cite{19}, it seems to be the appropriate
classical analog for the A.V. states \cite{20} \cite{21}. This picture is
derived also in the classical limit of the IBA-2 model \cite{22}. As it will
be shown in Sect. III, if the contra-rotation is considered slow with respect
to the intrinsic motion of the nucleons, and the resulting time-dependent 
wave function is properly quantized, then a semiclassical description of the
A.V. modes in the framework of the cranking model becomes possible.
\\ \indent
Along the lines sketched above, Sect. II contains the treatment of the pairing
interaction by using the HFB trial wave functions as SO(2n) coherent states.
As a counterpart, in Appendix the sp(4,R) structure of the single-particle
hamiltonian is emphasized, and closed formulas for the matrix elements of the
time-reversal operator are given. This algebraic formulation is intended to
allow a direct link between the present A.V. cranking model and the 
models based on boson expansions as IBA-2. \\ \indent
In this framework one can also do a further accurate treatment of the pairing
rotation and vibration modes in the rotating nuclei by global group
theoretical methods, giving new insight on the recently predicted pair
transfer \cite{23}. \\ \indent
The details of the proposed formalism describing semiclassically the A.V. 
states in realistic nuclei will be presented in Sect. III. Finally, in
Sect. IV the parameters are fixed, and some numerical results concerning the
ground-state rotational band and the A.V. modes in rare-earths nuclei are
discussed. \\[.5cm]
{\bf II. The Ground State Band} \\[.5cm] \indent
In the C.O. approximation for the single-particle part of the hamiltonian,
the writing of the pair operator $P^\dagger$ is difficult because the
basis eigenfunctions $\Psi_{n_1n_2n_3}$ (eq. A.8) are not time-reversal 
(complex conjugation) invariant due to the cranking term, and their 
complex conjugates $\Psi_{n_1n_2n_3}^*$ are not eigenfunctions due to the
anisotropy. In this case the usual pair operator $P^\dagger$, constructed by
the pairing of the time-reversed states, will be replaced by a more general one,
given by:
\begin{equation}
P^\dagger = \sum_{\alpha \beta} \langle \beta \vert \hat{K} \vert \alpha \rangle 
c^\dagger_\alpha c^\dagger_\beta = \sum_{a,b, \sigma} 
s_\sigma q_{ba} c^\dagger_{a \sigma} c^\dagger_{b - \sigma} ~~,
\end{equation}
where $a \equiv (n_1,n_2,n_3)$, $\alpha \equiv (a, \sigma)$, 
 $\vert \alpha \rangle = \Psi_{n_1n_2n_3} \vert \sigma \rangle$,   
$\sigma= \pm 1/2$ is the z-axis spin component, $s_\sigma = (-1)^{ \frac{1}{2}-
\sigma} $, $\hat{K}$ is the time-reversal operator, and $q_{ba} \equiv
\langle \Psi_b \vert \hat{K} \vert \Psi_a \rangle = \langle \Psi_b \vert
\Psi_a^* \rangle$. This expression originates in the commonly accepted form of a
separable interaction term in the particle-particle (pp) channel, $H_{int}=
- \chi P^\dagger P$. Here $P^\dagger$ is written in a basis independent form 
by using a single-particle operator $Q$ as $P^\dagger = \sum_{\alpha, \beta} 
\langle \beta \vert Q \vert \alpha \rangle c^\dagger_\alpha c^\dagger_\beta$
(ref. \cite{24} p. 180). If $Q=\hat{K} r^l Y^*_{lm}$ then the usual form of
the multipole pairing operator is recovered, and in particular, when $l=0$, this
reduces up to a constant factor to (1). \\ \indent
The ground state wave function $\vert Z \rangle$ for the proton or neutron
system is variationally obtainable, looking for the minimum of the energy $E=
\langle Z \vert H \vert Z \rangle/ \langle Z \vert Z \rangle$
with the constraint
\begin{equation}
\langle \hat{N} \rangle = \frac{ \langle Z \vert \hat{N} \vert Z \rangle }{
\langle Z \vert Z \rangle } = N_{part}
\end{equation}
imposed by a fixed mean-value $N_{part}$ of the particle number operator 
$\hat{N} = \sum_{a \sigma} c^\dagger_{a \sigma} c_{a \sigma}$. The many-body
hamiltonian $H$ is defined by
\begin{equation}
H = \sum_{a \sigma} \epsilon_a c^\dagger_{a \sigma} c_{a \sigma} - \frac{G}{4}
P^\dagger P
\end{equation}
where $G$ is the pairing constant and $\epsilon_a$ are dependent on the 
$\Omega_i$ frequencies (A.6,7) through
\begin{equation}
\epsilon_{n_1n_2n_3} = \sum_{i=1}^3  \Omega_i (n_i + \frac{1}{2})~~.
\end{equation}
If $\omega=0$ the matrix $q \equiv [q_{ab}]$ is diagonal, and the state $\vert Z 
\rangle \vert_{ \omega =0}$ has a BCS form:
\begin{equation}
\vert {\rm BCS} \rangle = \Pi_a \frac{1}{\sqrt{1+z_az_a^*}} e^{ \frac{1}{2} z_a P^\dagger_a}
\vert 0_p \rangle~~,~~
P^\dagger_a = \sum_\sigma  s_\sigma c^\dagger_{a \sigma} c_{a - \sigma}~.
\end{equation}
When $\omega >0$ the degeneracy of the energies $\epsilon_a$ changes, the matrix $q$
is no more diagonal, and HFB trial functions must be used. For simplicity it is
useful to treat these functions as elements of the K\"ahler manifold
\cite{25} of coherent states for the SO(2$n$) group:
\begin{equation}
\vert Z \rangle = \frac{1}{( {\rm det} \zeta )^{1/4}} e^{ \frac{1}{2} 
\sum_{\alpha, \beta=1}^n z^*_{\alpha \beta} c^\dagger_\alpha c^\dagger_\beta } 
\vert 0_p \rangle
\end{equation}
$$
\zeta = 1+ZZ^\dagger~~,~~Z^\dagger= (Z^T)^*
$$
parameterized by the antisymmetric complex matrix $Z=[z_{\alpha \beta}]$, and 
having the bare vacuum $\vert 0_p \rangle$ as the highest weight vector \cite{26}
\cite{27}. The integer $n$ specifying SO(2$n$) represents the number of single-particle
states $\alpha = (n_1,n_2,n_3, \sigma)$ under consideration. In this
formulation the calculus of the mean values 
$\langle Z \vert c^\dagger_\alpha c_\alpha \vert Z \rangle$,
$\langle Z \vert c^\dagger_\alpha c^\dagger_\beta  \vert Z \rangle$  and
$\langle Z \vert c^\dagger_\alpha c^\dagger_\beta c_\gamma c_\delta \vert Z \rangle$  
from the energy function $E$ is reduced to some derivatives of $\sqrt{ \rm det \zeta}$
with respect to the coordinates $z_{\alpha \beta}$.  Denoting by $\Delta = G \langle 
Z \vert P^\dagger \vert Z \rangle /2$, $\epsilon_{\alpha \beta}= \epsilon_a \delta_{ab}
\delta_{\sigma \sigma'}$, $Q_{\beta \alpha} = s_\sigma q_{ba}
\delta_{\sigma' - \sigma}$,
$\alpha \equiv (a, \sigma)$, $\beta \equiv (b, \sigma')$, $Tr(A)= \sum_\alpha A_{\alpha
\alpha}$, the expectation values $ \langle Z \vert \sum_{\alpha \beta} \epsilon_{\alpha 
\beta} c^\dagger_\alpha c_\beta \vert Z \rangle$, $\langle Z \vert P^\dagger \vert Z \rangle$ 
and $G \langle Z \vert P^\dagger P \vert Z \rangle /4$ are, respectively $Tr[ \epsilon (
1- \zeta^{-1})^T]$, $Tr[Q^T \zeta^{-1} Z]$ and
\begin{equation}
\frac{ \vert \Delta \vert^2 }{2} + \frac{G}{2} Tr[ QQ^\dagger -Q \zeta^{-1}
Q^\dagger - Q \zeta^{-1} Q^\dagger Z^\dagger \zeta^{-1} Z]~~.
\end{equation}
If in the last formula only the first term $\vert \Delta \vert^2 /2$ is retained, and the 
constraint (2) is accounted by introducing the Fermi level $\lambda$, then the 
variational equations $\partial (E- \lambda \langle \hat{N} \rangle )/ \partial z^*_{ \alpha 
\beta}=0$ can be explicitly written in a matrix form as
\begin{equation}
\Delta^* ZQZ-(fZ+Zf)+ \Delta Q^*=0~~,
\end{equation}
where $f=\epsilon- \lambda \hat{1}$. This equation can be further simplified
because the matrix $f$ is a constant in the spin space, and $Q$ factorizes
as $q \otimes \hat{s}$, with the matrix $q=[q_{ab}]$ real, symmetric, and
$\hat{s}_{ \sigma \sigma'} = (-1)^{\frac{1}{2} - \sigma}
\delta_{ \sigma' - \sigma}$. Thus, by choosing $Z$ as a direct product
$Z= \hat{t} \otimes \hat{s}$
(reduction from SO(2n) to USp(n) coherent states \cite{28} ) with $t$ real
and symmetric, and denoting by $trA \equiv \sum_a A_{aa}$, the equations (8),
(2) become
\begin{equation}
\hat{t}q \hat{t}+ \frac{1}{ \Delta} (f \hat{t} + \hat{t} f) = q
\end{equation}
\begin{equation}
2 tr[ \hat{t}^2(1+ \hat{t}^2)^{-1}]= N_{part}~~.
\end{equation}
Once we know the $\hat{t}$ matrix the ground state problem is solved, but the
description of the quasiparticle excitations, based on the calculus of
the matrices $U$ and $V$ from the HFB transformation
\begin{equation}
\tilde{c}^\dagger_\alpha = \sum_\beta U_{ \beta \alpha} c^\dagger_\beta +
V_{ \beta \alpha} c_\beta
\end{equation}   
is not completely achieved. The correspondence between the HFB and coherent states formulations 
becomes more evident by noticing that (8) follows from the defining equation
of $U$ and $V$ at the stationary point \cite{24} p. 613, \cite{29} :
\begin{equation}
\Delta^* U^\dagger Q U^* -(U^\dagger f V^* - V^\dagger f^T U^*) - \Delta V^\dagger Q^* V^* =0
\end{equation}
if $U$ is non singular and $Z=VU^{-1}$. \\ \indent
The system (9),(10) can be solved separately for protons and neutrons, obtaining
an $\omega$ - dependent matrix $\hat{t}$. To construct the intrinsic states of the
ground state rotational band
by using the functions $\vert Z_p (\omega_p) \rangle$, $\vert Z_n (\omega_n)
\rangle$ determined by the corresponding matrices $\hat{t}$, we suppose the
interaction between protons and neutrons being strong enough to prevent any
relative motion of their deformed mean-fields. Consequently, the angular
frequencies $\omega_p$ and $\omega_n$ are equal. For a state
with a given spin $I$, their common value $\omega_I$ is obtained from the
semiclassical condition ( \cite{24} p. 130) :
\begin{equation}
\langle Z_p \vert L_{ xp} \vert Z_p \rangle_{( \omega_I)} +    
\langle Z_n \vert L_{ xn} \vert Z_n \rangle_{( \omega_I)} = \sqrt{I (I+1)} ~~, I=0,2,4,...
\end{equation}
imposed on the expectation values of the many-body angular momentum operators $L_{xp}$,
$L_{xn}$. The rotational energy in the laboratory frame is the sum between the
proton and neutron terms, 
\begin{equation}
{\cal E}_I = {\cal E}_p ( \omega_I) + {\cal E}_n ( \omega_I)
\end{equation}
both defined by: ${\cal E}_\tau ( \omega) = E_\tau ( \omega) - E_\tau(0)$,
\begin{equation}
E_\tau ( \omega) =   \langle Z_\tau \vert H \vert Z_\tau \rangle_{( \omega)} +    
\omega \langle Z_\tau \vert L_{ x \tau} \vert Z_\tau \rangle_{( \omega)}~~, \tau=p,n. 
\end{equation}
As a function of $\omega$, $E_\tau$ will be interpolated in applications by using the 
following relation 
\begin{equation}
E_\tau ( \omega) = E_{0 \tau} + A_\tau \vert \omega \vert + \frac{ I_\tau \omega^2}{2}~~.
\end{equation}
Here $E_{0 \tau}$ represents the constant intrinsic energy and $I_\tau$ is interpreted as the
dynamical moment of inertia associated
with the classical free rotation around the x-axis. \\[.5cm]
{\bf III. The Angle Vibrations} \\[.5cm] \indent
For an elastic force acting between the proton and the neutron ellipsoids \cite{17}, at
the classical level a potential energy term\footnote{ this term can be also derived by
expanding in $\Phi$ the mean value $-
\chi \langle g \vert \hat{Q}_p \cdot \hat{Q}_n \vert g
\rangle $ of the microscopic quadrupole
p-n interaction \cite{34} taken on the laboratory functions
$\vert g \rangle$, (23).} $C \Phi^2/2$ depending on the angle
$\Phi$ formed by the symmetry (z) axes of the ellipsoids must be added to $E=E_p ( \omega_p) +
E_n ( \omega_n)$. Considering only rotations around the x-axis, the Lagrange function is,
up to a total time-derivative, given by
\begin{equation}
{\cal L} = \frac{ I_p \omega_p^2}{2} + \frac{ I_n \omega_n^2}{2} - \frac{C \Phi^2}{2}~~, 
\omega_p = \dot{ \phi}_p~,\omega_n = \dot{ \phi}_n~~.
\end{equation}
If new coordinates $\{ \Phi, \phi \}$ are introduced by the transformation
\begin{equation}
\Phi = \phi_p - \phi_n~~,~~ \phi= \frac{ I_p \phi_p + I_n \phi_n}{I_p+I_n}~~,
\end{equation}
then ${\cal L}$ can be written as a sum between a rotational and a vibrational part:
\begin{equation}
{\cal L} = \frac{ I_p+I_n}{2} \dot{ \phi }^2 + \frac{ I_pI_n}{I_p+I_n} \frac{ \dot{ \Phi}^2}{2} -
\frac{ C \Phi^2}{2} ~~.
\end{equation}
In these coordinates the classical Lagrange equations have a simple solution for the motion of the
proton and neutron mean-fields, representing an angular vibration with the frequency 
$\Omega= \sqrt{ C (I_p+I_n)/I_pI_n }$ superposed over an uniform rotation with the angular velocity
$\omega_r$:
\begin{equation}
\phi_p(t)= \omega_r t + a_p \sin \Omega t ~~,~~  \phi_n(t)= \omega_r t- a_n \sin \Omega t~~,
\end{equation}
\begin{equation}
\omega_p(t)= \omega_r + A_p \cos \Omega t ~~,~~  \omega_n(t)= \omega_r - A_n \cos \Omega t~~,
\end{equation}
where
\begin{equation}
A_p= \Omega a_p = \frac{I_n}{I_p+I_n} \omega_0 ~~,~~
A_n= \Omega a_n = \frac{I_p}{I_p+I_n} \omega_0~~.
\end{equation}
If the angle vibration takes place adiabatically, recalling the results of the cranking calculations 
presented in Sect. II, the wave functions of the ellipsoid $\tau=p,n$, in the intrinsic system and 
in the laboratory frame, are $\vert Z_\tau \rangle_{ ( \dot{ \phi}_\tau )} $  and
$\exp( - i \phi_\tau L_{x \tau} ) \vert Z_\tau \rangle_{ ( \dot{ \phi}_\tau )} $, respectively.
Considering only the case $\omega_r =0$, for the whole nucleus a periodic
time-dependent wave function $\vert g \rangle_{(t)}$ can now be obtained:
\begin{equation}
\vert g \rangle_{(t)} = e^{ - i \phi_p L_{x p} - i \phi_n L_{x n}} 
\vert Z_p \rangle_{ ( \omega_p )} \vert Z_n \rangle_{ ( \omega_n )} ~~.
\end{equation}
This function, subject to a method of quantization, will be used to give an
approximation to the  A.V. stationary state. The quantization problem for
the time-dependent solutions given by the various semiclassical
methods employed in the microscopic models \cite{30} \cite{31} and also for
some of the classical dynamics relevant in the description of the collective
modes in nuclei \cite{32} \cite{33} was much discussed, even if an unified
picture is not yet established. As it was remarked in \cite{31}, the wave
functions generated via the time-dependent variational principle from trial
functions with fixed norm are determined only up to a time-dependent phase
factor. Supposing this to be true also for $\vert g \rangle_{(t)}$
given above, the factor will be found by requiring an optimal time-evolution
of the function $\vert \psi \rangle_{(t)} = a_{(t)} e^{ i \theta (t) } \vert g
\rangle_{ (t)}$ relatively to the total microscopic hamiltonian H \cite{34}.
More precisely, the integral
\begin{equation}
{\cal I}_\psi=
\int dt \langle \psi \vert i \partial_t- {\rm H} \vert \psi \rangle 
\end{equation}
must be stationary at infinitesimal variations of the functions $a$ and
$\theta$, and
\begin{equation}
\delta_{ a, \theta} {\cal I}_\psi =0~~.
\end{equation}
A first step towards finding explicitly the stationary wave function is
made by retaining from the solution $\vert \psi \rangle$ of this equation
only the part invariant to the transformations of $\vert g \rangle$
by a time-dependent phase factor. This partial result,
$\vert \tilde{g} \rangle_{ (t)}$,
\begin{equation}
\vert \tilde{g} \rangle_{ (t)} = e^{ i \int_0^t dt' \langle g \vert i
\partial_{t'} \vert g \rangle } \vert g \rangle_{ (t)} 
\end{equation}
is a product between the periodic function $\vert g \rangle_{ (t)}$ and the
phase factor
$$
\exp( i \int_0^t dt' \langle g \vert i \partial_{t'} \vert g \rangle )~~,
$$
in general not periodic. Because the exact stationary solutions represent the
limit case when $\vert \tilde{g} \rangle$ is time-independent, a periodic
function $\vert \tilde{g} \rangle_{ (t)}$ is considered as a good
approximation for a constant one. This periodicity condition leads to a
constraint on the phase \cite{31}:
\begin{equation}
\int_0^T dt'  \langle g \vert i \partial_{t'} \vert g \rangle = 2 \pi n~,
~T= \frac{2 \pi}{ \Omega}~,
\end{equation}
which gives for each $n=1,2,...$ the amplitude $\omega_0$ in eq. (22),
undetermined up to now. \\
\indent
Finally, the stationary wave function $\vert \Omega \rangle$ is obtained
by projecting out the time
independent Fourier component of  $\vert \tilde{g} \rangle_{ (t) }$:
\begin{equation}
\vert \Omega \rangle = \frac{1}{T} \int_0^T dt \vert \tilde{g} \rangle_{ (t)}  ~~.      
\end{equation}
For applications to nuclei with a large number of particles,
$\vert g \rangle_{ (t)}$ is chosen as a product of four factors: 
$ \exp( - i \phi_p L_{x p}^v ) \vert Z_p \rangle_{ ( \omega_p )}$,
$ \exp( - i \phi_n L_{x n}^v ) \vert Z_n \rangle_{ ( \omega_n  )}$, and
$ \exp( - i \phi_p L_{x p}^c ) \vert \underline{ \alpha} _p
\rangle_{ ( \omega_p )}$,
$ \exp( - i \phi_n L_{x n}^c ) \vert \underline{ \alpha} _n \rangle_{
( \omega_n )}$, corresponding to the valence protons, valence neutrons, and
core protons, core neutrons, respectively. Here $\vert Z \rangle $ are the
HFB functions previously specified, and $\vert \underline{ \alpha} \rangle
\equiv \vert \alpha_1, \alpha_2, ..., \alpha_{n_c} \rangle$ are the Slater
determinants formed with $n_c$ single-particle states $\alpha_k$, $k=1,n_c$,
occupied by the core nucleons. Both functions are periodic time-dependent
through the oscillating frequencies $\omega_p (t)$, $\omega_n (t)$. The
expectation values required in the quantization formula (27) are given by
\begin{equation}
\langle Z_\tau \vert L^v_{x \tau} \vert Z_\tau \rangle = 2 tr[\hat{l}_x
\hat{t}^2 (1+ \hat{t}^2)^{-1}]~~,
\end{equation}
\begin{equation}
\langle \underline{\alpha}_\tau \vert L^c_{x \tau} \vert \underline{
\alpha}_\tau \rangle = 2 tr[\hat{l}_x]~~,
\end{equation}
 \begin{equation}
\langle Z_\tau \vert \partial_\omega \vert Z_\tau \rangle = \frac{1}{2}
tr \{ [ \hat{t}, \partial_\omega \hat{t}] (1+ \hat{t}^2)^{-1} \}~~,
\end{equation}
\begin{equation}
\langle \underline{\alpha}_\tau \vert \partial_\omega \vert \underline{
\alpha}_\tau \rangle = 0~~,
\end{equation}
where $\hat{l}_x$ is the matrix of the single particle angular momentum
operator $l_x$. \\ \indent In the following only the A.V. state with $n=1$
will be considered, approximating the quantized phase of the function $\vert
\tilde{g} \rangle_{ (t)}$ as $ \int_0^t dt'  \langle g \vert i \partial_{t'}
\vert g \rangle \approx \Omega t$. Thus, the state (28) becomes 
\begin{equation}
\vert \Omega \rangle = \frac{1}{T} \int_0^T dt e^{ i \Omega t} \vert g
\rangle_{ (t)}  ~~.
\end{equation}
Its norm 
\begin{equation}
{\cal N}^2 = \langle \Omega \vert \Omega \rangle = \frac{1}{T^2} \int_0^T
dt \int_0^T dt' e^{-i \Omega (t-t')}
\langle g(t) \vert g(t') \rangle 
\end{equation}
will be calculated assuming small angle oscillation amplitudes $a_p$, $a_n$
in eq. (20). In this case $\langle g(t) \vert g(t') \rangle$ has a simple
expression
\begin{equation}
\langle g(t) \vert g(t') \rangle= \Pi_{ \tau=p,n} w_\tau
( \omega_\tau(t)- \omega_\tau(t') )
W_\tau ( \omega_\tau(t)- \omega_\tau(t') )~~,
\end{equation}
where 
\begin{equation}
w_\tau ( \omega) \equiv \langle Z_\tau ( \omega) \vert Z_\tau (0) \rangle =
\frac{ {\rm det} [1+ \hat{t}_{( \omega)} ( \hat{t}_0')^\dagger]}{
\sqrt{ {\rm det} [1+ \hat{t}^2_{( \omega)}] {\rm det} [1+ \hat{t}_0'(
\hat{t}_0')^\dagger]}}~~,
\end{equation} 
\begin{equation}
W_\tau ( \omega) \equiv \langle \underline{ \alpha}_\tau ( \omega) \vert
\underline{ \alpha} _\tau (0) \rangle =
[{\rm det} (q^{ov} )^\dagger ]^2
\end{equation}
\begin{equation}
\hat{t}'_0= q^{ov \dagger} \hat{t}_{(0)} q^{ov *} ~~,
\end{equation}
and $q^{ov}$ is the overlap matrix $q^{ov}_{ab} = \langle \Psi_a (0) \vert
\Psi_b ( \omega) \rangle$ (A.10). \\ \indent
The state $\vert \Omega \rangle$ breaks the symmetries of the microscopic
hamiltonian because the projection of the particle and angular momentum
quantum numbers is approximate. In this form it can be used to evaluate only
the total strength
\begin{equation}
B(M1; 0^+ \rightarrow \Omega ) = \frac{ \vert \langle \Omega \vert {\cal M}
(M1) \vert 0 \rangle \vert^2 }{ \langle \Omega \vert \Omega \rangle  }
\end{equation}
of the magnetic transition from the $0^+$ ground state $\vert 0 \rangle \equiv
\vert g (0) \rangle$ to the A.V. mode $\omega_r=0$, $n=1$. The required
matrix elements
\begin{equation}
\langle \Omega \vert {\cal M}_{iv} (M1) \vert 0 \rangle = \frac{1}{T}
\oint dt e^{- i \Omega t} \langle g(t) \vert {\cal M}_{iv} (M1) \vert 0
\rangle ~~,
\end{equation}
\begin{equation}
\langle \Omega \vert {\cal M}_{is} (M1) \vert 0 \rangle = \frac{1}{T}
\oint dt e^{- i \Omega t} \langle g(t) \vert {\cal M}_{is} (M1) \vert
0 \rangle ~~,
\end{equation}
for the isovector
\begin{equation}
{\cal M}_{iv} (M1) = \frac{ g_p-g_n}{2} \sqrt{ \frac{3}{8 \pi}}
(L_{xp}-L_{xn}) \mu_N
\end{equation}
and the isoscalar 
\begin{equation}
{\cal M}_{is} (M1) = \frac{ g_p+g_n}{2} \sqrt{ \frac{3}{8 \pi}}
(L_{xp}+L_{xn}) \mu_N
\end{equation}
components, respectively, of the transition operator ${\cal M}(M1) =
{\cal M}_{is}(M1) + {\cal M}_{iv} (M1)$ will be  calculated by using the
microscopic values $g_p=1$, $g_n=0$. In the same approximation of
small amplitudes as above, we have:
\begin{equation}
\langle g(t) \vert L_{xp} \vert 0 \rangle = w_n(\omega_n) W_n (\omega_n)
[ f_p ( \omega_p) + F_p (\omega_p)] ~~,
\end{equation} 
\begin{equation}
\langle g(t) \vert L_{xn} \vert 0 \rangle = w_p(\omega_p) W_p (\omega_p)
[ f_n ( \omega_n) + F_n (\omega_n)] ~~,
\end{equation} 
where
\begin{equation}
f_\tau (\omega) \equiv \langle Z_\tau ( \omega) \vert L^v_{x \tau} \vert
Z_\tau (0) \rangle =
2 tr[ \hat{l}_x^T \hat{t} \hat{t}_0'^\dagger (1+ \hat{t}
\hat{t}_0'^\dagger)^{-1}] ~~,
\end{equation}
\begin{equation}
F_\tau( \omega) \equiv \langle \underline{ \alpha}_\tau ( \omega)
\vert L^c_{x \tau} \vert \underline{ \alpha}_\tau (0) \rangle = 2 W_\tau
( \omega) tr[ \tilde{l}_x (q^{ov \dagger})^{-1}]
\end{equation}
\begin{equation}
(\tilde{l}_x)_{ab} = ( \Psi_a (\omega) \vert l_x \vert \Psi_b (0) )~~. 
\end{equation}
Because the matrices $\hat{t}$ are not known as analytical functions of $\omega$,
to avoid a numerical integration in (34), (40) and (41), the functions $w$,
$W$, $f$, and $F$ obtained numerically will be interpolated by polynomials
in $\omega$, and in the end all the integrals will be calculated analytically.
\\[.5cm]
{\bf IV. Numerical Results} \\[.5cm] \indent  
a) The ground state band \\

In order to obtain the energy levels of the ground state band in rare-earths
nuclei it is necessary to solve the system of equations (9), (10) numerically.
In practice, it turns out to be convenient to restrict the SO(2$n$) (HFB)
treatment to a small number (10-12) of valence nucleons, while the other
particles are considered in a HF state. Such a type of calculation was done for
$^{156}_{~64}$Gd, $^{158}_{~64}$Gd and $^{168}_{~68}$Er, considering 12, 12, 6
valence protons and 10,12,0 valence neutrons, distributed on 12 and 9
single-particle states, respectively. These sets of states were chosen to
contain the Fermi level at $\omega=0$, and also to have no crossings
between their energy levels and the levels from outside, when $\omega$
increases. The constants of the model $\omega_1$, $\omega_2$, $\omega_3$
(A.1) and $G$ (3) can be expressed in terms of two parameters: the mass
number $A$ and the deformation $\delta$. Using the standard formulas
\cite{35}, $\omega_i$, $i=1,3$ are separately given for protons and neutrons
by the equations
\begin{equation}
\omega_1=\omega_2 = \omega_s ( 1+ \frac{ \delta}{3} ) ~~,~~
\omega_3 = \omega_s(1- \frac{2}{3} \delta ) ~~,~~
\omega_1 \omega_2 \omega_3 = \omega_{s \tau}^3~~,
\end{equation}
with $\omega_{sp} = 39.8 A^{-1/3}$ MeV and $\omega_{sn}= 44.8 A^{-1/3}$ MeV; 
$\delta$ has the values determined by Lamm \cite{35} 
\begin{equation}
\delta_{^{156}Gd}=0.239~~,~~\delta_{^{158}Gd}=0.250~~,~~\delta_{^{168}Er}=0.273~~,
\end{equation}
to have an optimal fit of the single-particle energy levels, and $G=23/A$ MeV. 
\\ \indent
The calculated rotational energy levels (eq. (13),(14)) and the experimental
ones are presented in figure 1. For $^{156}$Gd, a set of slightly
different parameters: $\delta=0.237$, $\omega_{sp}= \omega_{sn} =
41 A^{-1/3}$ MeV, yield
almost the same levels. However, a much different spectrum is obtained if the 
ground-state wave function is a Slater determinant of C.O. single-particle 
states (the HF approximation). As it can be seen from figure 2, in a HFB
state the $\omega$-dependence of the angular momentum for the valence nucleons of
$^{156}$Gd  is completely different than the one obtained by a HF calculation
(the results for protons and neutrons in this case are practically the same).   
 The HFB curves show a "diamagnetic" behavior \cite{15} correlated with the
variation of the order parameters $\Delta_p$, $\Delta_n$ represented in figure 3.
The existence of the superfluid layer also leads to differences between the
kinematical and dynamical moments of inertia extracted by interpolation from
the numerical results (with $\delta=0.239$, $\omega_{sp}=39.8 A^{-1/3}$ MeV,
$\omega_{sn}=44.8 A^{-1/3}$ MeV) obtained for the expectation
values $L_\tau ( \omega)  \equiv \langle g \vert L_{ x \tau} \vert g \rangle$,
$\tau=p,n$ of the angular momenta:
\begin{equation}
L_p ( \omega) = 19.8 \omega + 20 \omega^2~~,~~L_n ( \omega) = 33.7 \omega + 6.2 \omega^2~~,
\end{equation}
and of the energy ${\cal E}_\tau ( \omega) = E_\tau ( \omega) - E_\tau (0)$,
\begin{equation}
{\cal E}_p ( \omega) = \frac{ 37.9}{2} \omega^2 + 0.31 \vert \omega \vert
~~,~~ {\cal E}_n ( \omega) = \frac{ 44.1}{2} \omega^2 + 0.54 \vert \omega \vert~~,
\end{equation}
( $[L_\tau]= \hbar$, $[{\cal E}_\tau]=$ MeV). Such differences
practically do not appear in a HF calculation, when:
\begin{equation}
L_p ( \omega) = 34.4 \omega ~~,~~ L_n ( \omega) = 43.2 \omega ~~,
\end{equation}
\begin{equation}
{\cal E}_p ( \omega) = \frac{ 34.2}{2} \omega^2 ~~,~~ {\cal E}_n ( \omega ) = 
\frac{ 43.2}{2} \omega^2 ~~.
\end{equation}
It is interesting to remark that in all the cases investigated, the solution
$\hat{t}$ of the system (9),(10) has the same symmetry as the matrix $q$, in
the sense that $q_{ab}=0$ implies $ \hat{t}_{ab}=0$. This feature strongly
reduces the number of unknowns and also suggests that only the non-zero
elements of $\hat{t}$ are the important degrees of freedom in a further
time-dependent description of the pairing vibrations in rotating nuclei
\cite{23}, \cite{36} \cite{37}.
\\

b) The angle vibrations of $^{156}$Gd \\

A direct numerical calculation shows that $\langle Z_\tau \vert \partial_\omega
\vert Z_\tau \rangle$, $\tau=p,n$, are very small, and consequently the formula
(27) becomes simpler:
\begin{equation}
\int_0^T dt [ \dot{ \phi}_p L_p ( \omega_p) + \dot{ \phi}_n L_n ( \omega_n) ]
=2 \pi ~~.
\end{equation}
After the replacement of $\dot{ \phi}_\tau$ and $L_\tau$ with their explicit
formulas (21) ($I_p$, $I_n$ are given by (52)) and (51), one obtains
$\omega_0= \sqrt{ 2 \Omega /12.93}$, and the effective energy of the A.V.
state
\begin{equation}
E_{A.V.} = \frac{ I_p I_n}{I_p+I_n} \frac{ \omega_0^2}{2} = 1.57 \Omega ~~.
\end{equation}
The interpolation polynomials for the functions (36), (37), (46), (47) are 
explicitly
\begin{equation}
w_p ( \omega) = 1-6.2 \omega^2~~,~~ W_p ( \omega) = 1-3.47 \omega^2 ~~,
\end{equation} 
\begin{equation}
w_n ( \omega) = 1-8.7 \omega^2+27 \omega^4~~,~~ W_n ( \omega) = 1-3.6 \omega^2 ~~,
\end{equation} 
\begin{equation}
f_p ( \omega) = \omega(16.97-9.64 \omega)~~,~~ F_p ( \omega) =
\omega W_p( \omega ) (7.66-7.86 \omega)~~,
\end{equation} 
\begin{equation}
f_n( \omega) = 16.2  \omega ~~,~~ F_n ( \omega) = \omega W_n( \omega )
(8.52-9.5 \omega)~~.
\end{equation} 
The final results 
\begin{equation}
{\cal N}^2 ( \omega_0) = 2.7 \omega_0^2 - 17.56 \omega_0^4 + 60.65 \omega_0^6-
112.22 \omega_0^8+ 85.95 \omega_0^{10} ~~,
\end{equation}
\begin{equation}
\langle \Omega \vert {\cal M}_{iv} (M1) \vert 0 \rangle  = (2.12 \omega_0 -
4.73 \omega_0^3 + 4.0 \omega_0^5-
1.34 \omega_0^7+ 0.13 \omega_0^9) \mu_N ~~,
\end{equation}
\begin{equation}
\langle \Omega \vert {\cal M}_{is} (M1) \vert 0 \rangle  = (0.15 \omega_0 -
0.26 \omega_0^3 + 0.92 \omega_0^5-0.84 \omega_0^7+ 0.13 \omega_0^9) \mu_N ~~,
\end{equation}
contain the constant $\omega_0$, undetermined only by the parameters $\delta$
and $A$ appearing in the description of the ground state rotational band. In
order to obtain its value, the fit of one characteristic quantity for the
A.V. states with the experimental data  becomes necessary. An interesting
comment is that even before this fit, the $B(M1)_{iv}$ and $B(M1)_{is}$
strengths can be estimated in the limit of the small amplitude $\omega_0
\rightarrow 0 $:
\begin{equation}
B(M1)_{iv} \vert_{ \omega_0 \rightarrow 0} = 1.66 \mu_N^2 ~~,~~
B(M1)_{is} \vert_{ \omega_0 \rightarrow 0} = 0.009 \mu_N^2 ~~.
\end{equation}
Naturally the quantity which should be fitted is the elastic constant $C$, but
as its value is not precisely known, the fit will be on the energy $E_{A.V.}$,
by using the experimental value of the dominant line from the spectrum, 3.1
MeV. Then, all the other quantities, including $C$ can be calculated:
\begin{equation}
B(M1)_{iv} = 1.87 \mu_N^2~~,~~ B(M1)_{is}= 0.03 \mu_N^2
\end{equation}
\begin{equation}
a_n = 7.33~{\rm deg} ~~,~~a_p = 8.57~{\rm deg}
\end{equation}
\begin{equation}
C=79 ~{\rm MeV}~~,~~ \omega_0 = 0.55~{\rm MeV}
\end{equation}
As it was expected \cite{19} the value of $C$ given above is about four times
smaller than the value 311 MeV provided by the formulas of Palumbo
\cite{38}. The state $\vert \Omega \rangle$ proves to be excited from the
ground state mainly by the isovector component of the ${\cal M} (M1)$
operator, with a strength close to the experimental value for the total
strength of the $1^+$ fragments, $B(M1)_{iv,exp} = 2.3 \pm 0.5~\mu_N^2$
\cite{38} \cite{39}. The other results justify the approximation of small
oscillation amplitudes $a_p$, $a_n$. \\[.5cm] 
{\bf V. Summary and Conclusions} \\[.5cm] \indent
The starting point in this work was the treatment of the pairing interaction in 
deformed rotating nuclei. Assuming that the pairing term is separable in the pp
direction, and is determined only by a single-particle operator $Q$,
independently on basis, the expression (1) was obtained. In fact, with
$Q= \hat{K}$ this should also be found by rewriting the operator
$P^\dagger$ defined as usual in the spherical or Nilsson basis, if the sums
could be made infinite. Arguments supporting this choice are: the good
agreement with the experimental data obtained for the energy levels of the
ground-state rotational band, and the ease of the calculation, as almost
all quantities are known in closed analytical form. In applications the
only quantity which must be obtained numerically, the matrix $\hat{t}$,
was proved to have most of its elements 0, as they are in
the matrix $q$. This means in the case of a $12 \times 12$ matrix $\hat{t}$
that the number of unknowns $(\hat{t}_{ab}, \lambda)$ in the equation system
(9),(10) is reduced from 79 to only 18. Thus, by using the C.O. approximation
for the single-particle problem, by defining the pairing interaction using
the time-reversal operator, and the SO(2$n$) coherent states approach
to the many-body problem, we have an almost analytical model for the
treatment of the rotating superfluid
systems. As a byproduct, the overlap coefficients $q^{ov}$ (A.10) whose closed
formulas were not found in the literature, have their own importance in connection
to the representation theory of the Sp(4,R) group in the space of the
2-dimensional anisotropic harmonic oscillator states.  Another result was the
explicit form of the microscopic wave function for the A.V. states, including the
pairing correlations. This was derived within the cranking model, in the
adiabatic approximation for the motion of the proton and neutron deformed
potentials. In the quantization procedure, beside the standard results on the
energy spectrum, it was also touched the more delicate problem of the stationary
states. The prediction of the $B(M1)$ strengths is in a good agreement with the
experimental data, but of a greater importance appears to be the result on the
characteristic of the model represented by the limit value of $B(M1)_{iv}$ (64)
which is in the range of the experimental values for rare-earths nuclei.
This is an argument towards the physical relevance of the model in the A.V.
description, making it interesting for further studies. \\[.5cm]
{\bf Appendix:} The cranked oscillator \\[.5cm] \indent
In the cranking model for the anisotropic harmonic oscillator potential, the 
single-particle part of the hamiltonian has the form 
$$
h_{co} = \sum_{k=1}^3 \hbar \omega_k ( a^\dagger_k a_k + \frac{1}{2}) - 
\omega l_x ~~~~~~~~~~~~~(A.1)
$$
where
$$
a^\dagger_k = \sqrt{ \frac{ m \omega_k}{2 \hbar}} (x_k - \frac{i}{m \omega_k} p_k)
~~~~~~~~~~~~~~~~~~(A.2)
$$
$$
l_x= i \hbar [S (a_2^\dagger a_3-a_2a^\dagger_3)+D(a_2^\dagger
a_3^\dagger -a_2a_3)]~~~~(A.3)
$$
$$
S= - \frac{ \omega_2+ \omega_3}{ 2 \sqrt{ \omega_2 \omega_3} } ~~,~~
D= \frac{ \omega_3 - \omega_2}{ 2 \sqrt{ \omega_2 \omega_3}}
~~~~~~~~~~~~(A.4)
$$
This can be diagonalized exactly by a canonical transformation to new boson 
operators $A^\dagger_k$, $A_k$,
$$
A^\dagger_1=a^\dagger_1~~,~~A^\dagger_k = \sum_{j=2,3} x_{jk}
a_j^\dagger + y_{jk} a_j
~~~~~~~~~~(A.5)
$$
which satisfy, beside the canonical commutation relations, the equation
$$
[ h_{co}, A^\dagger_k] = \hbar \Omega_k A^\dagger_k~~~~~~~~~~~~~~~~~~~~~~~~
(A.6)
$$
$$
\Omega_1 = \omega_1~,~\Omega_{2,3}^2= \frac{ \omega_2^2 +
\omega_3^2 }{ 2} + \omega^2 \pm \frac{1}{2} \sqrt{ (\omega_2^2 -
\omega_3^2)^2 + 8 \omega^2 (\omega_2^2 + \omega_3^2) }~~~(A.7)
$$
It can be shown that in general, for the transformation (A.5) the scalar
product (overlap) between the elements of the set of functions 
$$
\Psi_{n_1n_2n_3} = \Pi_{k=1}^3 \frac{ (A^\dagger_k )^{n_k}}{
\sqrt{ n_k !}} \Psi_{000}~~,
~~A_k \Psi_{000}=0~~~~~~~~~~~(A.8)
$$
with elements from the set
$$
\Phi_{n_1n_2n_3} = \Pi_{k=1}^3 \frac{ (a^\dagger_k )^{n_k}}{ \sqrt{ n_k !}}
\Phi_{000}~~,
~~a_k \Phi_{000}=0
~~~~~~~~~~~~~(A.9)
$$
both obtained by acting with the operators $A_k^\dagger$, and $a_k^\dagger$ on
the vacuum states $\Psi_{000}$ and $\Phi_{000}$, respectively, is given by
$$
q^{ov}_{m_1m_2m_3, n_1n_2n_3} = ( \Phi_{m_1m_2m_3} \vert \Psi_{n_1n_2n_3} )
~~~~~~~~~~~~~~~~~~~(A.10)
$$
$$
= \delta_{m_1n_1} \sqrt{ n_2!n_3! m_2!m_3!} \sum_{k_2=0}^{n_2}
\sum_{k_3=0}^{n_3}
\sum_{j_1+j_2+l_1+l_2=k_2} \sum_{j_1'+j_2'+l_1'+l_2'=k_3} 
$$
$$  \times
R^{2 k_2 n_2}_{j_1j_2l_1l_2} R^{3k_3n_3}_{j_1'j_2'l_1'l_2'} \frac{ 
q^0_{m_2+j_1+j_1'-j_2-j_2',m_3+l_1+l_1'-l_2-l_2'}}{
\sqrt{( m_2+j_1+j_1'-j_2-j_2')!(m_3+l_1+l_1'-l_2-l_2')!}}~~.
$$
The coefficient $q^0_{n_2n_3} \equiv ( \Phi_{0n_2n_3} \vert \Psi_{000} )$
has non-zero values only if $n_2$, $n_3$ are both even or odd, and can be
calculated easily in a harmonic oscillator representation for the operators
$a_k^\dagger$, $a_k$. If $a_k = ( \xi_k + \partial/\partial \xi_k)/ \sqrt{2}$
then:
 $$
\Phi_{0n_2n_3} = \Pi_{k=2,3} \frac{ e^{ - \xi_k^2/2}}{\sqrt{ 2^{n_k} n_k!
\sqrt{ \pi} }} H_{n_k} ( \xi_k)
~~~~~~~~~~~~~~~(A.11)
$$
$$
\Psi_{000} =
( \frac{ u_2u_3}{\pi^2} )^{1/4}
e^{ - \frac{u_2}{2} \xi_2^2 - \frac{u_3}{2} \xi_3^2 - i s \xi_2 \xi_3}
~~~~~~~~~~~~~~~~(A.12)
$$
where 
$$
u_k = 1/[1+ 2 \sum_{j=2}^3 ( \vert y_{kj} \vert^2 - x_{kj} y^*_{kj})]
~~~~~~~~~~~~~~~~~(A.13)
$$
$$
s= 2 {\rm Im}[ \frac{ x_{22} y_{23} - x_{23} y_{22} }{(x_{23} -
y_{23})(x_{32} -y_{32})-(x_{22} - y_{22})(x_{33}-y_{33})} ]
~~~~~~(A.14)
$$
and 
$$
q^0_{n_2n_3} = (-i)^{I_{n_2}}
\sqrt{
\frac{ n_2!n_3! \sqrt{u_2u_3} }{ (u_2+1)(u_3+ \eta +1) 2^{n_2+n_3-2} } }
\sum_{k=I_{n_2}}^{n_2}
\frac{ (-1)^{ \frac{k-I_{n_2}}{2} } ( \frac{1-u_2}{1+u_2} )^{
\frac{n_2-k}{2}} ( \frac{s}{1+u_2})^k  }{ k! ( \frac{n_2-k}{2})!}
$$
$$
\times
\sum_{m= \frac{k+I_k}{2} }^{ \frac{n_3+k}{2}}
\frac{ (2m)! (-1)^{( \frac{n_3+k}{2} -m)} }{ m! (2m-k)!
( \frac{n_3+k}{2} - m)! ( \frac{s+ \eta +1}{2} )^m }~~~~~~~~~~~(A.15)
$$
with 
$$
I_k = \frac{ 1- (-1)^k}{2} ~~,~~ \eta =\frac{s^2}{u_2+1}~~.
$$
The factors $R^{jkn}_{j_1j_2l_1l_2}$ are defined by the expansion
$$
(A^\dagger_j)^n = n! \sum_{k=0}^n \sum_{j_1+j_2+l_1+l_2=k}
R^{jkn}_{j_1j_2l_1l_2}
a_2^{j_1}(a_2^\dagger)^{j_2} a_3^{l_1} (a_3^\dagger)^{l_2} ~~~~~~(A.16)
$$
and are, explicitly
$$
R^{jkn}_{j_1j_2l_1l_2} = \sum_{j_3+l_3=n-k} (1-I_{j_3})(1-I_{l_3})
(-1)^{ \frac{n-k}{2}}
$$
$$ \times
\frac{ (y_{2j})^{j_1+j_3/2} (x_{2j})^{j_2+j_3/2} (y_{3j})^{l_1+l_3/2}
(x_{3j})^{l_2+l_3/2} }{2^{(n-k)/2} j_1!j_2! ( \frac{j_3}{2} ) !
( \frac{l_3}{2})! l_1! l_2! } ~~~~~~~~(A.17)
$$
The time-reversal operator $\hat{K}$ acts on $\Psi_{n_1n_2n_3}$
through complex conjugation, and to obtain the matrix $q$, defined as
$q_{m_1m_2m_3,n_1n_2n_3} = ( \Psi_{m_1m_2m_3} \vert \hat{K} \vert
\Psi_{n_1n_2n_3})$ it is enough to use the expressions derived above for
the case when  $(A^\dagger_j)^*$, $(A_j)^*$ are written in terms of
$A^\dagger_j$, $A_j$. Because in this case
$$
(A_1^\dagger)^* = A_1^\dagger~~,~~ ( A^\dagger_k)^* = \sum_{j=2,3}
\alpha_{jk} A^\dagger_j + \beta_{jk} A_j ~~~~~~~~~(A.18)
$$
by choosing the phases such that $x_{jk}^* =
(-1)^j x_{jk}$, $y^*_{jk} = (-1)^j y_{jk}$, $j=2,3$,
$$
\alpha_{jk}^* = x_{2j}x_{2k}-y_{2j}y_{2k}+x_{3k}x_{3j}-y_{3j}y_{3k}
~~~~~~~~~~~~~(A.19)
$$
$$
\beta_{jk} = x_{2j} y^*_{2k} -x^*_{2k}y_{2j}+x_{3j}y^*_{3k}-x^*_{3k}y_{3j}
~~~~~~~~~~~~~(A.20)
$$
are real numbers, we have $s=0$, and $q^0_{n_2n_3}$ are also real,
non-zero only for $n_2$, $n_3$ even, having the simple expression
$$
q^0_{n_2n_3} = \frac{1}{ ( \frac{n_2}{2})! ( \frac{n_3}{2})!} 
( \frac{1-u_2}{1+u_2})^{\frac{n_2}{2}} ( \frac{1-u_3}{1+u_3})^{\frac{n_3}{2}}
\sqrt{ \frac{n_2!n_3! \sqrt{u_2u_3} }{2^{n_2+n_3-2} (1+u_2)(1+u_3)} }~.
~~(A.21)
$$
It is interesting to remark that excepting for the term
$\hbar \omega_1 (a_1^\dagger a_1 + \frac{1}{2}) $ the C.O. hamiltonian is a
linear combination of the representation operators for the sp(4,R) Lie algebra
$\{ p_jp_k,p_jq_k+q_kp_j,q_jq_k \}$ $j,k=2,3$ \cite{40} \cite{41} in the
Hilbert space ${\cal H}$ of the anisotropic harmonic oscillator states, and
the transformation (A.5) corresponds to a unitary representation operator
${\cal U}$ of the group Sp(4,R) in ${\cal H}$:
$$
A^\dagger_k ( \omega) = {\cal U}( \omega)  a^\dagger_k {\cal U}^{-1}( \omega)
~~~~~~~~~~~~(A.22)
$$
$$
\Psi_{n_1n_2n_3} ( \omega) = {\cal U} ( \omega) \Phi_{n_1n_2n_3}
~~~~~~~~~~(A.23)
$$  
whose matrix elements are the coefficients $q^{ov}$. Moreover, the operators
(A.2) are connected to the boson operators for the isotropic harmonic
oscillator $b^\dagger_k = \sqrt{m \omega_s/2 \hbar} (x_k - i p_k/m \omega_s)$
by the
transformation $a^\dagger_k = {\cal V}( \theta_k )  b^\dagger_k {\cal V}^{-1}
( \theta_k) $. Here ${\cal V} ( \theta_k )$ is an unitary operator generated
by the element $( (b_k^\dagger)^2-b_k^2)/2$ of the representation for sp(2,R)
Lie algebra in the Hilbert space of the harmonic oscillator states:
$$
{\cal V} ( \theta_k) = e^{ \theta_k [(b_k^\dagger)^2-b_k^2]/2 }
~~~~~~~~~~~~~~~~(A.24)
$$
$$
a^\dagger_k = \cosh \theta_k b^\dagger_k - \sinh \theta_k b_k~~~~~~~~~~(A.25)
$$
$$
\cosh \theta_k = \frac{1}{2} ( \sqrt{ \frac{ \omega_k}{ \omega_s} } +
\sqrt{ \frac{ \omega_s}{ \omega_k} } )~~~~~~~~~~(A.26)
$$
Denoting by $\phi_0 ( \xi_k)$ the vacuum state for the $b_k$ bosons, and by
$\phi_{n_k} = (n_k!)^{-1/2} (b_k^\dagger)^{n_k} \phi_0( \xi_k)$, the
"deformed" vacuum $\Phi_{000}$ becomes
$$
\Phi_{000} = \Pi_{k=1}^3 {\cal V}( \theta_k) \phi_0 ( \xi_k)=
\Pi_{k=1}^3 \sum_{n_k=0}^\infty
\frac{ \sqrt{ (2n_k)!} ( \tanh \theta_k )^{n_k} }{
2^{n_k} n_k! \sqrt{ \cosh \theta_k } } \phi_{2 n_k} ~~~(A.27)
$$
  
\vskip1cm
{\bf Figure Captions} \\[.5cm] \noindent
Fig. 1. The experimental and calculated energy levels of the ground state band
for $^{156}$Gd, $^{158}$Gd, $^{168}$Er. For $^{156}$Gd it is shown also the
HF result, neglecting the pairing interactions. \\
Fig. 2. The angular momentum as a function of the rotation frequency for the 
considered valence nucleons of $^{156}$Gd ($\delta=0.237$, $\omega_{sp} =
\omega_{sn}=41 A^{-1/3}$ MeV). \\
Fig. 3. The order parameters $\Delta_p$, $\Delta_n$ for the valence nucleons 
of $^{156}$Gd ($\delta=0.237$, $\omega_{sp} = \omega_{sn}=41 A^{-1/3}$ MeV). 
\end{document}